\documentclass[letterpaper]{article} % DO NOT CHANGE THIS
\usepackage{aaai24}
\usepackage{times}  % DO NOT CHANGE THIS
\usepackage{helvet}  % DO NOT CHANGE THIS
\usepackage{courier}  % DO NOT CHANGE THIS
\usepackage[hyphens]{url}  % DO NOT CHANGE THIS
\usepackage{graphicx} % DO NOT CHANGE THIS
\usepackage{natbib}  % DO NOT CHANGE THIS AND DO NOT ADD ANY OPTIONS TO IT
\usepackage{caption} % DO NOT CHANGE THIS AND DO NOT ADD ANY OPTIONS TO IT
\usepackage{algorithm}
\usepackage{algorithmic}

\usepackage{amsmath}
\DeclareMathOperator{\sign}{sign}
\usepackage{multirow} %KD: is this okay??
\usepackage{subfigure}
\usepackage{dblfloatfix}

\usepackage{newfloat}
\usepackage{listings}
\DeclareCaptionStyle{ruled}{labelfont=normalfont,labelsep=colon,strut=off} % DO NOT CHANGE THIS
\floatstyle{ruled}
\newfloat{listing}{tb}{lst}{}
\floatname{listing}{Listing}
\title{Analyzing the Impact of Tax Credits on Households in Simulated Economic Systems with Learning Agents}

\author {
    Jialin Dong\textsuperscript{\rm 1}\footnote{This work was done when Jialin Dong was a summer intern at JP Morgan AI Research.},
    Kshama Dwarakanath\textsuperscript{\rm 2},
    Svitlana Vyetrenko\textsuperscript{\rm 2}
}
\affiliations {
    \textsuperscript{\rm 1}University of California at Los Angeles, Los Angeles, CA.\\
    \textsuperscript{\rm 2}JP Morgan AI Research, Palo Alto, CA.\\
    jialind@g.ucla.edu, kshama.dwarakanath@jpmorgan.com, svitlana.s.vyetrenko@jpmchase.com
}
\usepackage{bibentry}
\begin{document}

\maketitle

\begin{abstract}
In economic modeling, there has been an increasing investigation into multi-agent simulators. Nevertheless, state-of-the-art studies establish the model based on reinforcement learning (RL) exclusively for specific agent categories, e.g., households, firms, or the government. 
It lacks concerns over the resulting adaptation of other pivotal agents, thereby disregarding the complex interactions within a real-world economic system. 
% It lacks concerns over other pivotal agents, i.e., the central bank, which sets the interest rate, thereby disregarding the complex interactions within a real-world economic ecosystem. 
Furthermore, we pay attention to the vital role of the government policy in distributing tax credits. Instead of uniform distribution considered in state-of-the-art, it requires a well-designed strategy to reduce disparities among households and improve social welfare. To address these limitations, we propose an expansive multi-agent economic model comprising reinforcement learning agents of numerous types. 
Additionally, our research comprehensively explores the impact of tax credit allocation on household behavior and captures the spectrum of spending patterns that can be observed across diverse households. Further, we propose an innovative government policy to distribute tax credits, strategically leveraging insights from tax credit spending patterns. Simulation results illustrate the efficacy of the proposed government strategy in ameliorating inequalities across households.
\end{abstract}

\section{Introduction}

In the domain of economic modeling, the spotlight has shifted towards multi-agent simulators. Recent papers like \cite{bucsoniu2010multi, nowe2012game, 2004.13332, curry2022analyzing} have contributed insights into the dynamics among households, firms, and the government.
Furthermore, a line of literature, e.g., \cite{dosi2019more, axtell2022agent, haldane2019drawing, liu2022welfare, hill2021solving, chen2021deep}, illuminates the impressive results of reinforcement learning (RL) within economic modeling. A pivotal contribution by \cite{sinitskaya2015macroeconomies} proposes an economic model characterized by uniform consumer-workers and firms responsible for price and wage setting. Moreover, the work \cite{curry2022analyzing} introduces a more complicated model including consumers, firms, and the government, and learns an equilibrium in open real-business-cycle models.

Nonetheless, the research mentioned above does possess certain limitations that need to be addressed. Firstly, the papers \cite{dosi2019more, axtell2022agent, haldane2019drawing, liu2022welfare,sinitskaya2015macroeconomies,hill2021solving} confine the utilization of reinforcement learning to a select category of agents, thereby lacking the modeling of other adaptive agents (such as the government or the central bank or both). This issue inadvertently neglects the complexities of real-world economic systems. 
Secondly, state-of-the-art such as \cite{curry2022analyzing} uniformly distribute tax credits and neglect the significance of government intervention in maintaining social welfare through tax collection and redistribution. Even though a recent paper \cite{liu2022welfare} proposed an efficient algorithm to achieve social welfare maximization and
competitive equilibrium simultaneously in the economic system, it only considers two types of agents, i.e., agents and planners.

To design an optimal tax credit distribution strategy, it's important to study the tax credit spending patterns across various household types. In light of a comprehensive research report by JP Morgan Chase \cite{JPMC}, household liquidity, defined as the ratio of savings to consumption, emerges as a powerful predictor of consumption behavior when tax credits are distributed to households. The report shows that low-liquidity households tend to allocate a larger proportion of their tax credits towards consumption, as opposed to higher-liquidity households. Notably, the spending patterns across various household types remain unexplored in multi-agent-based economic systems.
Understanding these spending patterns is crucial for policymakers and economists aiming to formulate targeted and effective tax credit distribution policies. By studying the consumption behaviors of low-liquidity households, valuable insights can be gained into their financial challenges and needs. Such insights can inform the development of an effective credit distribution strategy, ultimately promoting welfare and economic growth.
% Furthermore, bridging the gap in knowledge regarding spending behaviors within multi-agent-based economic systems could pave the way for innovative research and policy initiatives, fostering a more nuanced understanding of how tax credits impact diverse households. By acknowledging and addressing these intricacies, policymakers can create policies that not only stimulate economic activity but also enhance the overall well-being of households across different liquidity levels.

Our paper models the dynamics and objectives of households, firms, central bank and the government in a multi-agent simulator. By enabling learning based strategies for each of the various agent types, we aim to create a more realistic and nuanced representation of the economic system. By integrating the central bank's activities into our analysis, we capture the pivotal role it plays by adjusting interest rates to manage inflation and achieve GDP targets \cite{blinder2008central}. Moreover, we offer a comprehensive investigation of the impact of tax credit allocation on diverse household behavior. Based on this investigations, we design a government policy to facilitate an equitable distribution of tax credits to promote social welfare.

Our contributions are summarized as follows:
\begin{itemize}
\item We consider a multi-agent economic model that encompasses households, a firm, the central bank, and the government. Our implementation of this simulator builds upon a state-of-the-art simulator for financial markets \cite{byrd2019abides} and its extension \cite{amrouni2021abides}.
\item For an exploration of the spending behaviors of high-liquidity and low-liquidity households in response to tax credits, we introduce an innovative household dynamics model. This model incorporates a new action to capture distinct spending patterns. Simulation outcomes reveal that lower-liquidity households tend to spend a greater share of their tax credits, aligning with the findings observed in a JP Morgan Chase report \cite{JPMC}.

\item Leveraging the insights from the tax credit spending patterns, we propose a novel government strategy aimed at optimizing the distribution of tax credits to enhance overall social welfare. Our simulation results confirm the effectiveness of this proposed strategy, illustrating its potential to mitigate inequalities among diverse households.
\end{itemize}

\section{Economic model}\label{sec:econ_model}
	In this section, we present an economic model that involves households earning labor income from a firm and consuming goods, the firm producing goods using household labor, the government collecting income tax from households, and distributing tax credits to households \cite{krusell1998income,evans2005policy,benhabib2001monetary}. At the same time, the central bank sets the interest rate to achieve target inflation and improve the GDP \cite{kaplan2018monetary,svensson2020monetary}. Distinguished from state-of-the-art, every agent in our system uses reinforcement learning to adapt their strategies to maximize individual objectives. Thus, providing a more comprehensive understanding of the dynamics of the economic model in the real world.

 % \textcolor{blue}{Markov game since multiple RL agents. All finite horizon objectives!}
Our economic model with multiple reinforcement learning agents can be formalized as a Markov Game (MG) with each agent having partial observability of the global system state \cite{littman1994markov,hu1998multiagent}. 
% Markov decision process (MDP) \cite{littman1994markov} with multiple types of agents. 
The MG consists of finite-length episodes of $H$ time steps where each time step $t\in\{0,1,\cdots,H-1\}$ corresponds to one-quarter of economic simulation. We detail the model concerning each type of agent below. Agents' parameters, observations, actions, and rewards are summarized in Table \ref{table}.
	% \usepackage{multirow}

	\begin{table*}[t]
		\centering
		\begin{tabular}{|c|l|c|c|}
			\hline
			\textbf{Agent}                 & \multicolumn{1}{c|}{\textbf{Observation}}                                                                                                                                                                                                                                                                                                                                                                            & \textbf{Action}                                   & \textbf{Reward}                                     \\ \hline
			\multirow{3}{*}{Household $i$} & \begin{tabular}[c]{@{}l@{}}$(m_{t,i}, r_{t}, \tau_{t},\kappa_{t,i},p_{t},w_{t}
				
				%				 \textcolor{red}{,\eta_{t,i}}
				
				)$\\ $m_{t,i}$: saving\\ $r_t$: interest rate\\ $\tau_{t}$: tax rate\\ $\kappa_{t,i}$: tax credit\\ $p_{t}$: price of good set by firm\\ $w_{t}$: wage set by firm\\
				
				%			\textcolor{red}{$\eta_{t,i}$: the proportion of consumption paid by tax credits}
			\end{tabular} & \multirow{3}{*}{\begin{tabular}[c]{@{}l@{}}
				$(n_{t,i},c_{t,i}^{act})
				%				 \textcolor{red}{,\eta_{t,i}} 
				$\\$n_{t,i}$: labor hours \\ $c_{t,i}^{act}$: requested consumption 
			\end{tabular}} & \multirow{3}{*}{(\ref{reward_h})} \\ \cline{2-2}
			& \textbf{Parameter
				%				\textcolor{red}{(I am not confident with the names of the parameters are proper enough.)}
			}                                                                                                                                                                                                                                                                                                                                                                                             &                                                   &                                                     \\ \cline{2-2}
			& \begin{tabular}[c]{@{}l@{}}$(\gamma_i,\nu_i,\mu_i)$\\ $\gamma_i$: {parameter for isoelastic utility of consumption and savings}\\ $\nu_i$: {coefficient for disutility of labor}\\ $\mu_i$: {coefficient for isoelastic utility of savings} \end{tabular}                                                                                                                                                                                              &                                                   &                                                     \\ \hline
			\multirow{3}{*}{Firm}      & \begin{tabular}[c]{@{}l@{}}$({\epsilon_{t-1}}, Y_{t},{p_{t}, }w_{t},\sum_i n_{t,i}, \sum_i c_{t,i})$\\$\epsilon_{t-1}$: exogenous production factor\\ $\varepsilon_{t}$: production shock \\ $Y_{t}$: inventory of firm\\ $n_{t,i}$: labor hours of household $i$\\ $c_{t,i}$: realized consumption of household $i$\end{tabular}                                                                                                                                                       & \multirow{3}{*}{$(w_{t+1},p_{t+1})$}            & \multirow{3}{*}{(\ref{reward_F})} \\ \cline{2-2}
			& \textbf{Parameter}                                                                                                                                                                                                                                                                                                                                                                                             &                                                   &                                                     \\ \cline{2-2}
			& \begin{tabular}[c]{@{}l@{}}$(\alpha,\rho,\sigma,\chi)$\\ $\alpha$: efficiency of production using labor \\ $\rho$: auto-regression coefficient for production factor \\ $\sigma$: standard deviation of production shock\\$\chi$: coefficient for inventory risk\end{tabular}                                                                                                                                                                                                                                          &                                                   &                                                     \\ \hline
			Central Bank                   & \begin{tabular}[c]{@{}l@{}}$(p_{t-4}, p_t, y_{t})$\\  Calculates inflation $\pi_t$ using (\ref{inflation})\end{tabular}                                                                                                                                                                                                                                                                           & $r_{t+1}$                                             & (\ref{reward_cb})                 \\ \hline
			Government                     & \begin{tabular}[c]{@{}l@{}}$(\tau_t,\{\kappa_{t,i}\}_i,\sum_i\tau_tn_{t,i}w_t )$\\
   % $\tau_tn_{t,i}w_t $: tax revenue \\

   \end{tabular}                                                                                                                                                                                                                                            & $(\tau_{t+1},\{\kappa_{t+1,i}\}_i)$                 & (\ref{reward_g})                  \\ \hline
		\end{tabular}
		\caption{Summary of parameters, observations, actions, and rewards of agents in the economic model.}\label{table}
	\end{table*}
	
	\subsection{Households}
	At each time step $t$, household $i$ works for $n_{t,i}$ hours and attempts to consume $c^{act}_{t,i}$ units of the good produced by the firm. 
	The good's price is set by the firm as $p_{t}$, and an income tax rate is set by the government as $\tau_t$. Furthermore, the realized consumption of households depends on the available inventory of goods $Y_{t}$. The total demand for goods produced by the firm is given by $\sum_i c_{t, i}^{a c t}$. If there is insufficient supply to meet the demand, goods are rationed proportionally among households as follows:
	\begin{align*}
	c_{t, i}=\min \left\{c_{t,i}^{act}, Y_{t} \cdot \frac{c_{t, i}^{a c t}}{\sum_i c_{t, i}^{a c t}}\right\}.
	\end{align*}
	
	Household savings increase according to the interest rate $r_t$ set by the central bank. Additionally, consumption and work affect a household's savings $m_{t}$. Household $i$ receives labor income equal to $ n_{t,i} w_{t}$, where the firm pays a wage $w_{t}$.
	
	The cost of consumption for household $i$ is given by $p_{t} \cdot c_{t,i}$. Moreover, with a flat tax rate of $\tau_t$, households pay income tax amounting to $\tau_t \cdot  n_{t,i} w_{t}$. The government redistributes the tax revenue as tax credits $\kappa_{t, i}$ to ease household budgetary constraints.
 Considering all these factors, household savings change according to the following equation:
 \begin{equation}\label{saving}
     \begin{aligned}
         m_{t+1, i}=&\left(1+r_t\right) m_{t, i}+\left(n_{t, i} w_{t}-p_{t} c_{t, i}\right)\\
 &-{\tau_tn_{t,i}w_t}+\kappa_{t, i}.
     \end{aligned}
 \end{equation}
	
	Each household $i$ optimizes its labor hours and consumption to maximize utility \cite{evans2005policy}:
 \begin{align}
     \max _{\left\{n_{t, i}, c_{t, i}^{act}\right\}_{t=0}^{H-1}} \sum_{t=0}^{H-1} \beta_{i, \mathbf{H}}^t  u\left(c_{t, i}, n_{t, i}, m_{t+1, i} ; \gamma_{i}, \nu_i, \mu_i\right)\nonumber
 \end{align}
 where
	\begin{align}\label{reward_h}
	u(c, n, m ; \gamma, v, \mu)=\frac{c^{1-\gamma}}{1-\gamma}+\mu\sign(m)\frac{|m|^{1-\gamma}}{1-\gamma} -\nu n^2
	\end{align}
	with $ \beta_{i, \mathbf{H}}$ being the household's discount factor.
	The utility function is a sum of isoelastic utility from consumption with parameter $\gamma_i$, isoelastic utility from savings with parameter $\gamma_i$ and coefficient $\mu_i$, and a quadratic disutility of work with coefficient $\nu_i$.
	
	\subsection{Firm}
	At each time step, the firm employs households to produce goods and earns money from consumed goods. At time step $t$, it determines a price $p_{t+1}$ for its goods and selects a wage $w_{t+1}$ to pay, which will come into effect at the next time step $t + 1$.
	
	Based on the total labor hours $\sum_i n_{t,i}$, the firm produces $y_{t}$ units of goods per the following Cobb-Douglas production function \cite{cobb1928theory}:
	\begin{equation}\label{production}
	y_{t}=\epsilon_{t}\left(\sum_i n_{t, i}\right)^{\alpha}
	\end{equation}
	where $0 \leq \alpha \leq 1$ characterizes the efficiency of production using labor\footnote{We pass the production function (\ref{production}) through a floor function to compute the integer number of goods produced.}. And, $\epsilon_{t}$ is the exogenous production factor which follows a log-autoregressive process with coefficient $0 \leq \rho \leq 1$ given by
	\begin{equation*}
	\epsilon_{t}=\left(\epsilon_{t-1}\right)^{\rho} \exp \left(\varepsilon_{t}\right),
	\end{equation*}
	which allows for an exponential increase with a shock $\varepsilon_{t}\sim\mathcal{N}(0,\sigma^2)$. 
 % The coefficient $0 \leq \rho \leq 1$ represents an autoregression coefficient.
	
	Furthermore, households purchase $ \sum_i c_{t,i}$ units of the goods produced by the firm. Consequently, the inventory of the firm changes as $Y_{t+1} = Y_{t} + y_{t} - \sum_i c_{t,i}$.
 The firm optimizes the prices of its goods and the wages of households to maximize profits, resulting in the following formulation:
 \begin{equation}\label{reward_F}
     \max_{\{w_{t}, p_{t}\}_{t=0}^{H-1}} \sum_{t=0}^{H-1} \beta_{ \mathbf{F}}^t\left(p_{t} \sum_ic_{t, i}-w_{t}\sum_in_{t, i}-\chi p_{t} Y_{t+1}\right),
 \end{equation}
 % \begin{equation}\label{reward_F}
 %     \begin{aligned}
 %         \max_{\{w_{t}, p_{t}\}_{t=0}^{H-1}} \sum_{t=0}^{H-1} \beta_{ \mathbf{F}}^t\Big(p_{t} \sum_ic_{t, i}-w_{t}&\sum_in_{t, i}\\
 % &-\chi p_{t} Y_{t+1}\Big),
 %     \end{aligned}
 % \end{equation}
	% \begin{align}\label{reward_F}
	% \max_{\{w_{t}, p_{t}\}_{t=0}^{H-1}} \sum_{t=0}^{H-1} \beta_{ \mathbf{F}}^t\Big(p_{t} \sum_ic_{t, i}-w_{t}&\sum_in_{t, i} \notag\\
 % &-\chi p_{t} Y_{t+1}\Big),
	% \end{align}
	where $\beta_{ \mathbf{F}}$ represents the firm's discount factor, $\chi>0$ is the regularization coefficient to restrict the firm from overly increasing the price to make profits without a subsequent increase in consumption. One can interpret this term as the risk of holding inventory as a result of producing much more than the expected consumption in the absence of the assumption of market clearing.
	\subsection{Central Bank}
	The central bank gathers data on annual price changes and the production of the firm at each time step. At time step $t$, the central bank sets the interest rate $r_{t+1}$, which will come into effect in the economic system at the next time step $t+1$. The annual inflation rate at time step $t$ is derived by the central bank based on the annual price changes and is calculated as:
	\begin{align}\label{inflation}
	\pi_t=\frac{ p_{t}}{ p_{t-4}}.
	\end{align}
	
	Additionally, the central bank evaluates the GDP using the production of the firm $ y_{t}$.
	The central bank optimizes its monetary policy strategy to achieve its inflation and GDP targets as follows \cite{svensson2020monetary,hinterlang2021optimal}:
	\begin{align}\label{reward_cb}
	\max _{\left\{r_t\right\}_{t=0}^{H-1}} \sum_{t=0}^{H-1} \beta_{\bf{CB}}^t\left(-\left(\pi_t-\pi^{\star}\right)^2+\lambda y_{t}^2\right),
	\end{align}
	where $\beta_{\mathbf{CB}}$ represents the central bank's discount factor, $\lambda>0$ is the weight given to productivity relative to the inflation target, and $\pi^{\star}$ is the desired inflation target. 
 % The central bank aims to maximize this objective function over an infinite time horizon.
	
	\subsection{Government}
   {At time step $t$, the government collects income taxes denoted by $\sum_i\tau_tn_{t,i}w_t$, and sets the tax rate $\tau_{t+1}$ that will be effective at the next time step $t+1$.}
	% The government sets income tax rates and collects tax revenue, denoted as $T( n_{t, i} w_{t}, \tau_t)$. 
 % At time step $t$, the government chooses an interest rate $\tau_{t+1}$ that will be effective at the next time step $t+1$. 
 At time step $t$, the government {distributes the collected tax revenue through tax credits $\kappa_{t+1,i}$} that will be distributed to household $i$ at the next time step $t+1$.
	
	The government optimizes its choice of tax rate and distribution of tax credits to maximize social welfare:
	\begin{align}\label{reward_g}
	\max _{\left\{\tau_{t}, \{\kappa_{t, i}\}_i\right\}_{t=0}^{H-1}} \sum_{t=0}^{H-1} \beta_{ \mathbf{G}}^t P\left(\tau_{t}, \lbrace\kappa_{t, i}\rbrace_i\right)
	%\left(\sum_{i}P\left(\tau_{t}, \kappa_{t, i}\right)\right),
	\end{align}
	where $\beta_{ \mathbf{G}}$ represents the government's discount factor, and $P\left(\tau_{t}, \lbrace\kappa_{t, i}\rbrace_i\right)$ represents the social welfare at time step $t$. The detailed definition of social welfare will be discussed in Section \ref{sec:gov}.
 
	% \textcolor{blue}{Can we remove the notation $T( n_{t, i} w_{t}, \tau_t)$ and change it to $\tau_tn_{t,i}w_t$ or $\sum_i\tau_tn_{t,i}w_t$ as appropriate throughout the paper?}

 \section{Tax Credit Study}\label{sec:3}
	In this section, we study the impact of tax credits on the dynamics of the economic system. 
	\subsection{Household consumption based on liquidity}\label{subsec:3.1}
	According to a research report by JP Morgan Chase \cite{JPMC}, households with lower liquidity tend to allocate a larger portion of their tax credits towards consumption, in contrast to those with higher liquidity. Here, liquidity is a measure of the cash buffer of the household measuring its savings relative to consumption spending. At time step $t$, the liquidity of household $i$ is defined as 
	\begin{align}\label{liquidity}
	l_{t,i} = \frac{m_{t,i}}{p_{t}c_{t,i}}.
	\end{align}
 To demonstrate the ability of our economic model to accurately capture this trend in real data, we introduce a new action for households that characterizes the source of spending for consumption. Let $\eta_{t,i}\in[0,1]$ denote the proportion of consumption spending that the household $i$ pays using tax credits. Given the finite nature of tax credits, we impose constraints on $\eta_{t,i}$ to ensure realistic spending. Specifically, we require that:
	\begin{align}\label{pay_constraint}
	p_{t}c_{t,i}\eta_{t,i}\leq \kappa_{t,i}.
	\end{align}

 To better understand how households with varying levels of liquidity utilize their tax credits, we propose an augmented utility function that accurately reflects the impact of this action on consumption choices.
 % To incorporate this crucial phenomenon into our economic model, we introduce a new action for households to characterize the pattern of spending tax credits.  
 % We propose a revised utility function that accurately reflects the impact of this action on consumption choices.
	% To facilitate a thorough examination of the effects of tax credits, we begin by introducing several key parameters that will play a crucial role in our study. 
	% To better understand how households with varying levels of liquidity utilize their tax credits, we introduce an action for each household, denoted as $\eta_{t,i}\in[0,1]$. This variable signifies the proportion of consumption that the household $i$ pays using tax credit. 
 % \textcolor{blue}{Remove $\sum_j$ for below constraints since we have a single firm.}
	% Next, we present our proposed utility function that effectively captures the impact of tax credits. 
 As tax credits often serve to stimulate consumption \cite{blundell2000labour, ellwood2000impact}, we shape the utility from spending tax credits in a manner akin to the utility of consumption as described in equation (\ref{reward_h}) as
	\begin{align}\label{utility}
	{f(\eta)} =\frac{(\eta c)^{1-\gamma}}{1-\gamma}.
	\end{align}
	The utility function (\ref{utility}) suggests that households can derive more benefits as they spend more tax credit, provided that $\eta$ obeys the constraint (\ref{pay_constraint}). 
 
 Leveraging the dynamics of $m_{t,i}$ as defined in (\ref{saving}), and the utility function $f(\eta)$ as described in (\ref{utility}), households maximize their overall utility via optimization {of} their labor hours, consumption, and the proportion of consumption spending paid by tax credits as
 \begin{align*}
     &\max _{\left\{{\eta_{t,i}},n_{t, i}, c_{t, i}^{act}\right\}_{ t=0}^{H-1}}\sum_{t=0}^{H-1} \beta_{i, \mathbf{H}}^t  u\left({\eta_{t,i}},c_{t, i}, n_{t, i}, m_{t+1, i};  {\gamma_{i}}, \nu_i, {\mu_i}\right)
 \end{align*}
 subject to $p_{t}c_{t,i}\eta_{t,i}\leq \kappa_{t,i}$ where \begin{equation}
     \begin{aligned}
        u( {\eta}, c, n, m; \gamma, \nu, \mu)=&\frac{c^{1-\gamma}}{1-\gamma}+\mu \sign(m)\frac{|m|^{1-\gamma}}{1-\gamma}\\
     &-\nu n^2+\frac{(\eta c)^{1-\gamma}}{1-\gamma}.
     \end{aligned}\nonumber
 \end{equation}
 % \begin{align}
 %     &\max _{\left\{{\eta_{t,i}},n_{t, i}, c_{t, i}^{act}\right\}_{ t=0}^{H-1}}\sum_{t=0}^{H-1} \beta_{i, \mathbf{H}}^t  u\left({\eta_{t,i}},c_{t, i}, n_{t, i}, m_{t+1, i};  {\gamma_{i}}, \nu_i, {\mu_i}\right)\nonumber\\
 %     &\textnormal{subject to }p_{t}c_{t,i}\eta_{t,i}\leq \kappa_{t,i}\nonumber\\
 %     &\textnormal{where }u( {\eta}, c, n, m; \gamma, v, \mu)=\frac{c^{1-\gamma}}{1-\gamma}+\mu \cdot\sign(m)\frac{|m|^{1-\gamma}}{1-\gamma}\nonumber\\
 %     &-\nu n^2+{f({\eta})}.\nonumber
 % \end{align}
 % where $u( {\eta}, c, n, m; \gamma, v, \mu)=\frac{c^{1-\gamma}}{1-\gamma}-v n^2+\mu \cdot\sign(m)\frac{|m|^{1-\gamma}}{1-\gamma}+{f({\eta})}$. 
 The simulation results presented in Section \ref{sec: result} demonstrate that low-liquidity households tend to allocate a larger portion of their tax credit towards consumption, in contrast to high-liquidity households. For a comprehensive analysis and more detailed simulation results, please refer to Section \ref{sec:results_prop}.
	\subsection{Distributing tax credits to improve social welfare}\label{sec:gov}
	To enhance social welfare, the distribution strategy of tax credits by the government is paramount. Specifically, focusing on aiding low-liquidity households can significantly contribute to their financial well-being. In this section, we present an innovative government model designed to optimize the allocation of tax credits to improve social welfare.
	
	% To begin with, we introduce parameters, states, and actions in the government model. 
 We consider the scenario where the government distributes all the collected income tax $\sum_i\tau_tn_{t,i}w_t$ from time step $t$ as tax credits to the different households at time step $t+1$. Mathematically, this can be expressed as $\sum_i \kappa_{t+1,i}=\sum_i\tau_tn_{t,i}w_t$.
	To enhance the financial well-being of households, the government's objective can be divided into two key components: firstly, improving overall household benefits, and secondly, allocating more substantial tax credits to low-liquidity households. Consequently, the government objective function (\ref{reward_g}) is defined as follows:
	\begin{align*}
 % \label{govern_r}
	\max _{\left\{\tau_{t}, \kappa_{t, i}\right\}_{t=0}^{H-1}} \sum_{t=0}^{H-1} \beta_{ \mathbf{G}}^t\left(\theta\cdot\mathcal{H}_t+
	\sum_{i}\frac{1}{l_{t,i}}\cdot {\kappa_{t+1,i}}\right),
	\end{align*}
	where $l_{t,i}$ (\ref{liquidity}) is the liquidity of household $i$ at time step $t$, $\theta\in(0,1)$ is a weighting coefficient for $\mathcal{H}_t$ which is the total reward across all households at time step $t$. The advantage of the proposed government model will be illustrated by the simulation results in Section \ref{sec:results_gov}.

	\section{Experimental Results}\label{sec: result}
  \begin{figure*}[!b]
		\centering
		\includegraphics[width=\linewidth]{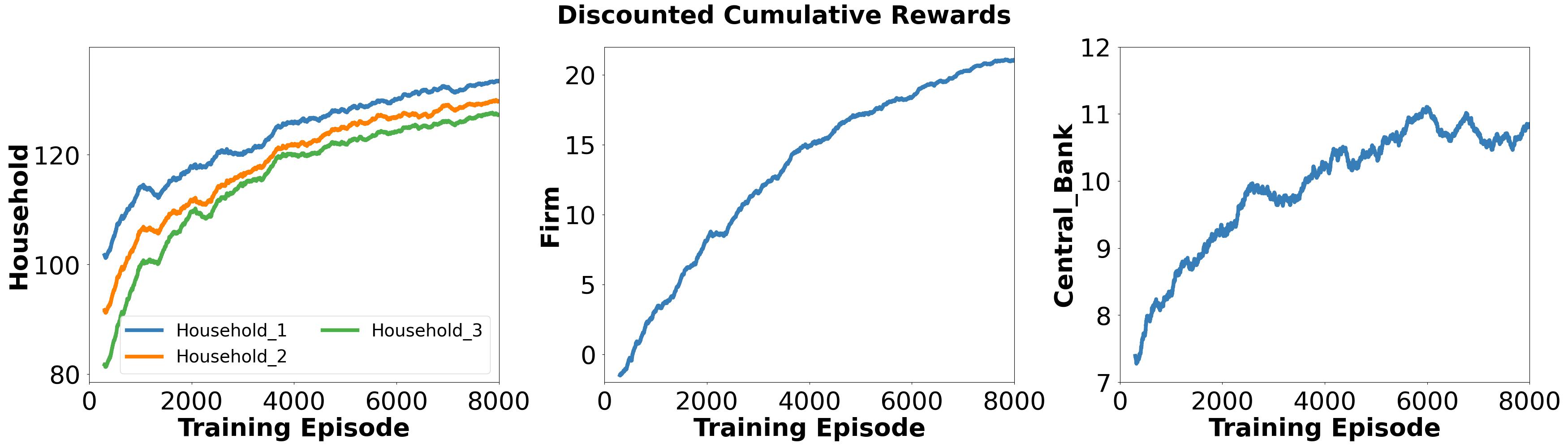}
		\caption{Training rewards without tax credits.}\label{fig:tax_credit}
	\end{figure*}
    \subsection{Multi-Agent Simulator}
	Our simulations are implemented based on a state-of-art multi-agent simulator called ABIDES \cite{byrd2019abides}, and its OpenAI Gym extension called ABIDES-gym \cite{amrouni2021abides}.
 % \footnote{https://github.com/jpmorganchase/abides-jpmc-public}. 
 It is a high-fidelity agent-based
	simulator used by practitioners and researchers {in the financial domain} \cite{vyetrenko2020get,dwarakanath2022equitable}. 
 We adapt the simulator to model interactions between the economic agents described in Section \ref{sec:econ_model} and enable multi-agent reinforcement learning using the Gym extension of the simulator.
 % The model is generated by simulating the interactions between a given set of agents, for which we can define the number, type, and strategy to simulate different economic systems. 
 % More details on the simulator can be found in the original ABIDES paper \cite{byrd2019abides}.

	\subsection{Learning details}
 %    \begin{figure*}[t]
	% 	\centering
	% 	\includegraphics[width=\linewidth]{fig/tc_0_train}
	% 	\caption{Training rewards without tax credits.}\label{fig:tax_credit}
	% \end{figure*}
	We describe the learning setup in more detail here.
	\begin{itemize}
		\item Households: We consider three households to demonstrate the efficacy of our models in this work although our setup is suitable for the use of many more. Each household's consumption choices range from $0$ to $24$ units of goods in increments of $6$ units. Their labor choices range from $0$ to $960$ hours per quarter, in increments of $240$ hours. The proportion of consumption spending paid by tax credits can vary between $0$ and $1$, in units of $0.25$, subject to the constraint described in (\ref{pay_constraint}). The disutility of labor coefficient $\nu_i = 0.5$ for all households. All households start with savings $m_{0,i}=0$.
		
		\item Firm: There is a single firm in the model. The firm's choices for good prices range from $188$ to $456$ dollars, in increments of $67$ dollars. The wage choices for the firm range from $7.25$ to $56.87$ dollars per hour, in increments of $12.405$ dollars. The firm has labor efficiency $\alpha=0.67$, auto-regression coefficient $\rho=0.97$, shock standard deviation $\sigma=0.1$, and inventory risk coefficient $\chi=0.1$. The starting value of the exogenous production factor is $\epsilon_0=1$ with starting inventory $Y_0=0$. 
  % The inventory coefficient $\chi$ in (\ref{reward_F}) is set to $0.1$. The technology shock factor is $_{t}=0.5$. 
  % \textcolor{blue}{Check value of all parameters!}
		
		\item Central Bank: The central bank in the model can set interest rates that range from $0.25\%$ to $5.7\%$, in increments of $1.375\%$. The target inflation $\pi^*$ is $1.02$ and the coefficient for productivity $\lambda=1$. 
		
		\item Government: The government chooses the income tax rate from a set of values: $0.10, 0.12, 0.22,$ $ 0.24, 0.32, 0.35, 0.37$. 
  % In our paper, the government distributes a tax credit of \$5000 per household per quarter. 
  The weighting coefficient $\theta$ for $\mathcal{H}_t$ in the government's reward is set as $0.1$. 
  % \textcolor{blue}{What are the choices for tax credit amounts?}
	\end{itemize}
All agents have discount factor $\beta_{i,\mathbf{H}}=\beta_{\mathbf{F}}=\beta_{\mathbf{CB}}=\beta_{\mathbf{CB}}=0.99$. To ease learning, we normalize agent rewards so that their objectives are given as follows:
	\begin{itemize}
		\item Household:  \[\sum_{t} \beta_{i, \mathbf{H}}^t  u (c_{t, i}, \frac{n_{t, i}}{\tilde{n}}, \frac{m_{t+1, i}}{{\tilde{n}}\cdot\sum_{j}{\tilde{w}}}; \gamma_{i}, \nu_i, \mu_i ), \]
		where $\tilde{n}=480$ is the median labor hours in households' action space, and $\tilde{w}=32.06$ is the median wage in the firm's action space.
		\item Firm: \begin{align*}
		  \sum_{t} \beta_{\mathbf{F}}^t&\left(\frac{p_{t}\sum_ic_{t,i}}{\tilde{p}\sum_k\tilde{c}}-\frac{w_{t}\sum_in_{t,i}}{\tilde{w}\sum_k\tilde{n}}
    -\chi \frac{p_{t} Y_{t+1}}{\sum_i\exp(1)\tilde{p}\tilde{n}}\right)
		\end{align*}
		where $\tilde{p}=322$ is the median price in the firm's action space, and $\tilde{c}=12$ is the median consumption in households' action space.
		\item Central bank:\[	 \sum_{t} \beta_{\bf{CB}}^t\left(-\left(\pi_t-\pi^{\star}\right)^2+\lambda\left( \frac{y_{t}}{ \tilde{y}}\right)^2\right),\]
		where $\tilde{y} = \left(\sum_i \tilde{n}\right)^{\alpha}$.
		\item Government: \begin{align*}
		\sum_{t} \beta_{ \mathbf{G}}^t\left(\theta\cdot\mathcal{H}_t^{\mathrm{norm}}+\sum_{i}\frac{1}{l_{t,i}}\cdot \frac{\kappa_{t+1,i}}{\sum_k{\tilde{p}}\cdot{\tilde{c}}}\right)
		\end{align*}
		where $\mathcal{H}_t^{\mathrm{norm}}$ is the sum of normalized rewards at $t$ across all households.
	\end{itemize}
  % \textcolor{blue}{Need to put discount factors for all agents.}
	We use the Proximal Policy Optimization algorithm in the RLlib package to learn strategies for all reinforcement learning agents.

	\begin{figure*}[t]
		\centering
		\subfigure{%
			{\includegraphics[width=0.85\columnwidth]{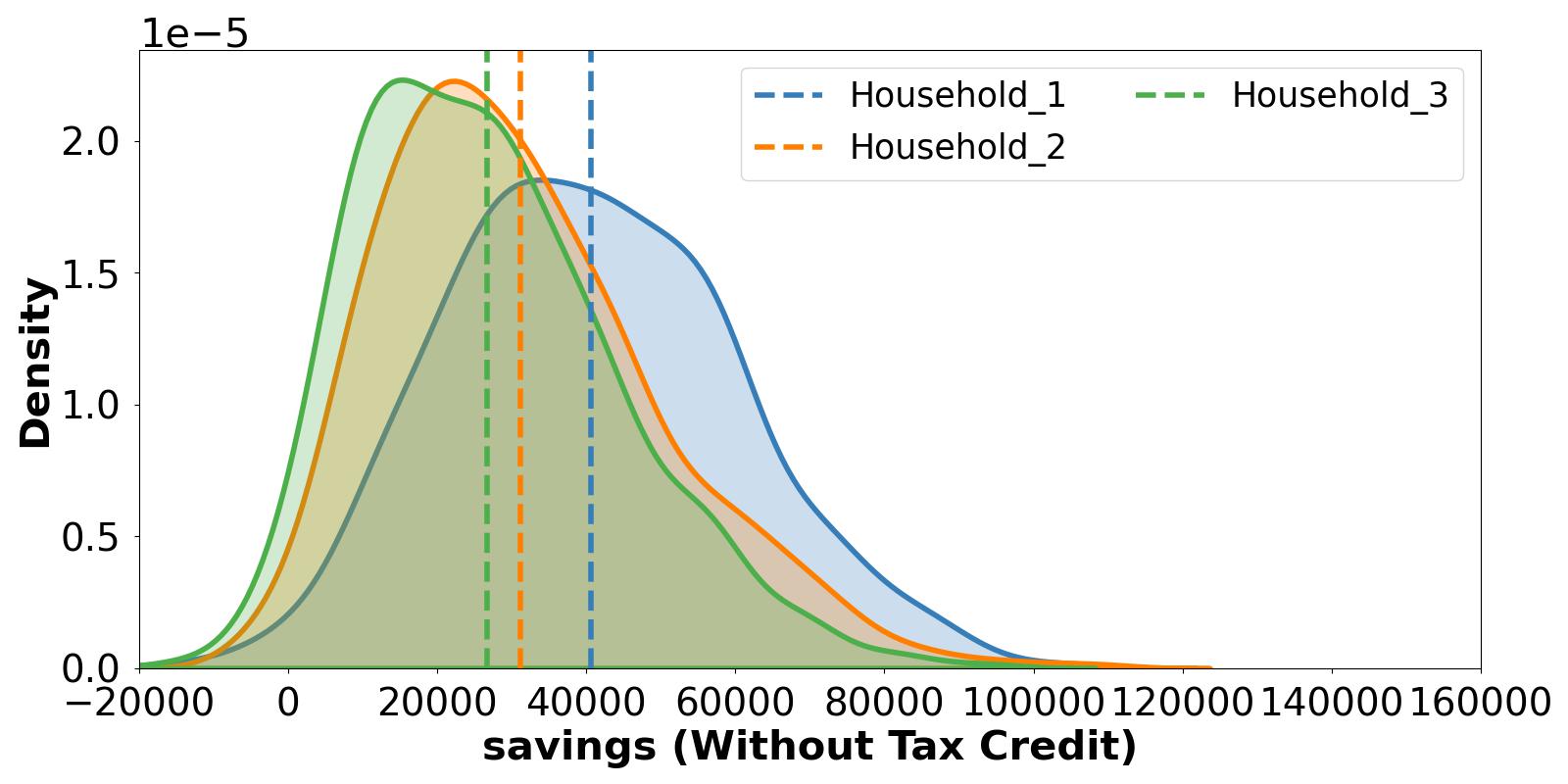}}%
			
		}
        \subfigure{%
			{\includegraphics[width=0.85\columnwidth]{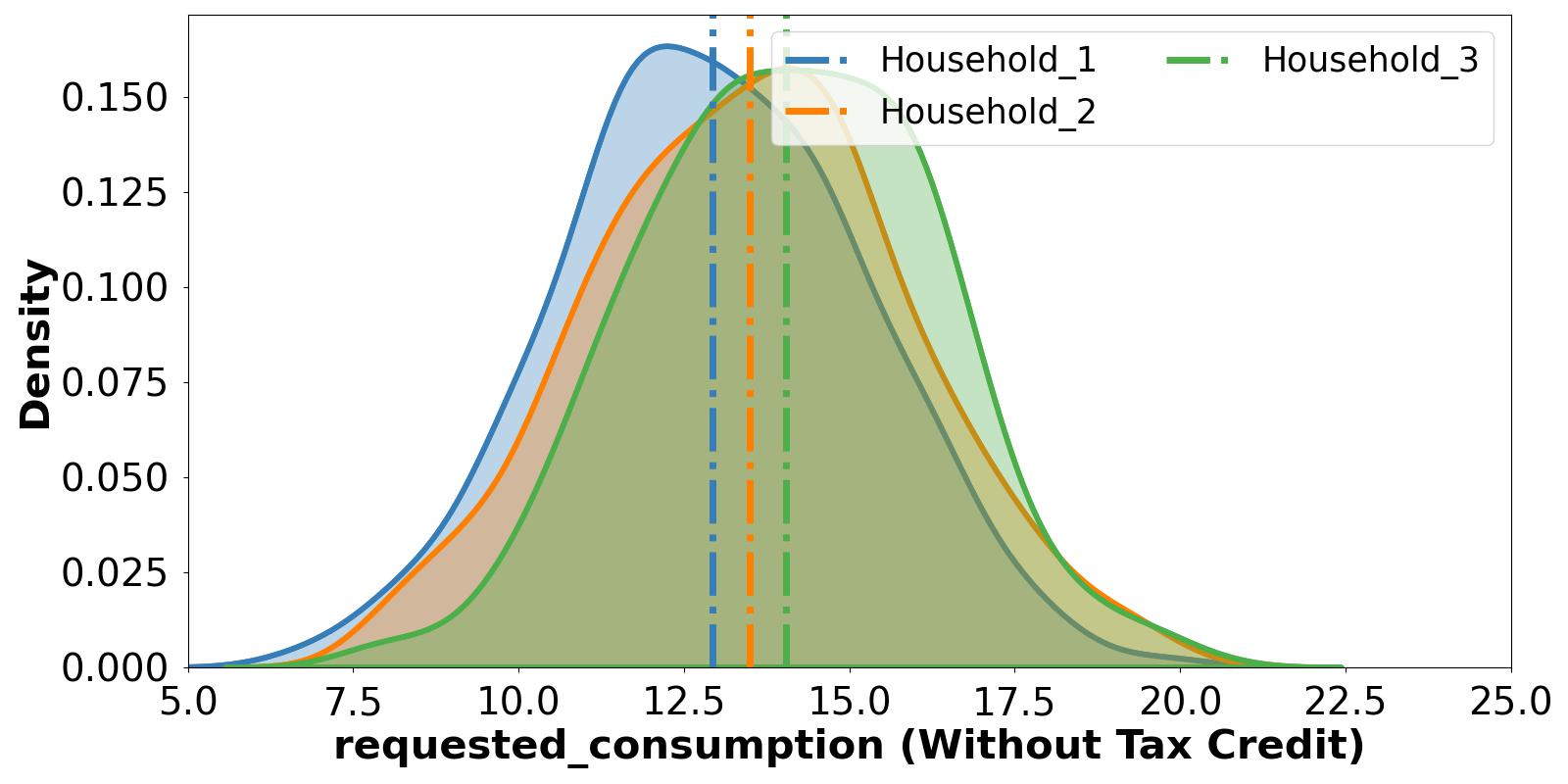}}%
			
		}
		\subfigure{%
			{\includegraphics[width=0.85\columnwidth]{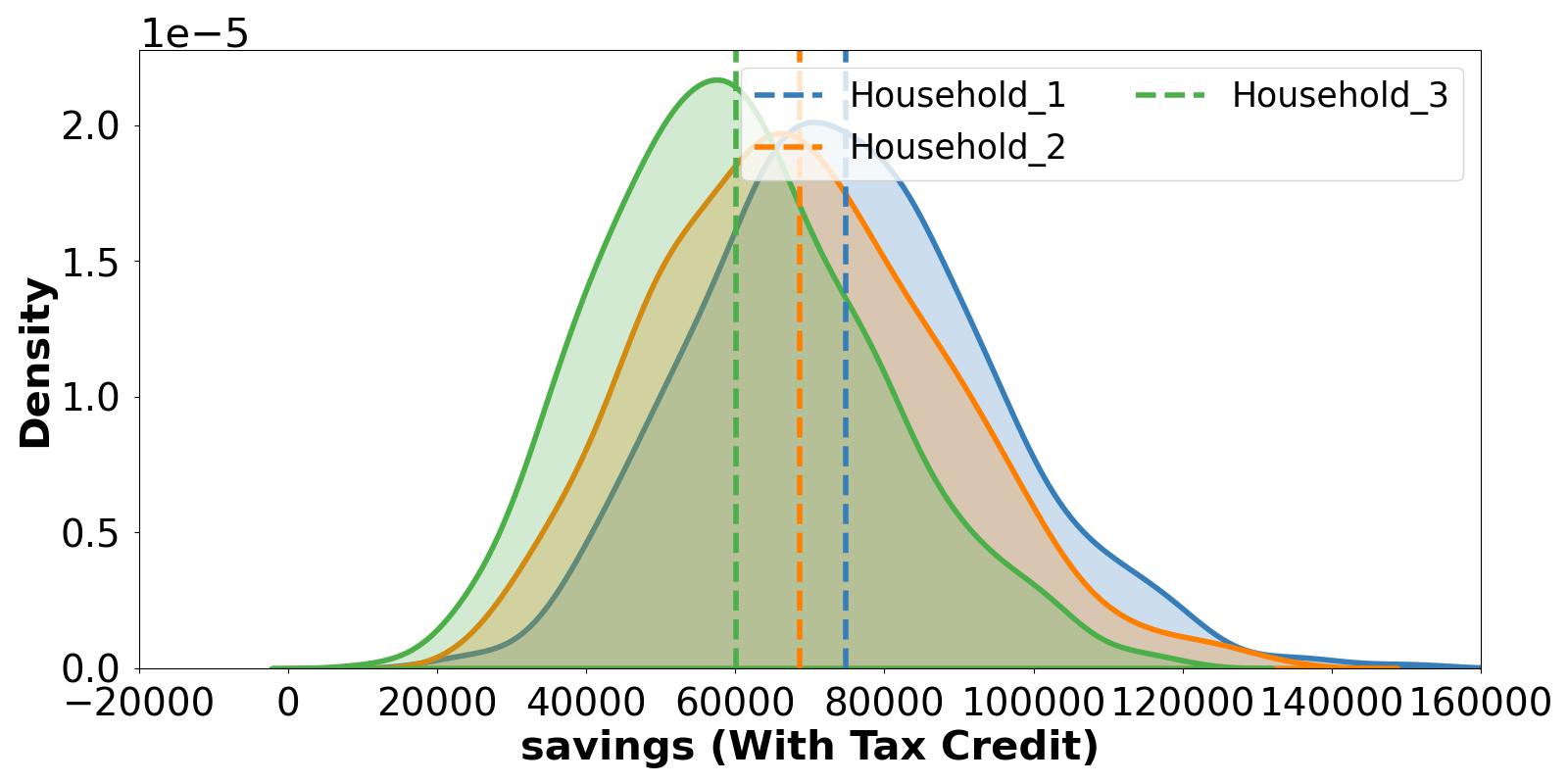}}%
			
		}
  		\subfigure{%
			{\includegraphics[width=0.85\columnwidth]{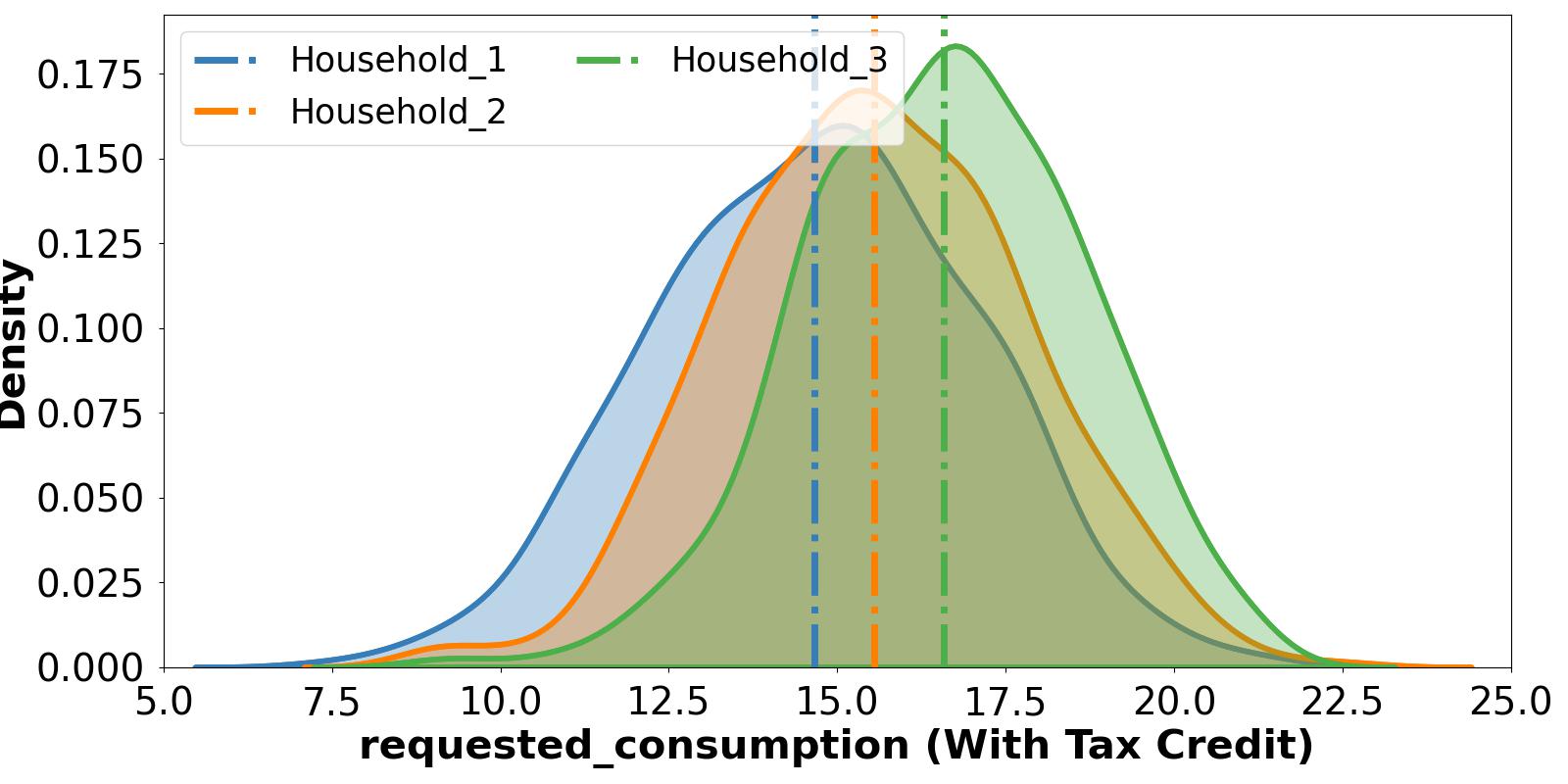}}%
			
		}
		% \subfigure{%
		% 	{\includegraphics[width=0.4\columnwidth]{fig/Household_labor_hours_dist_0}}%
			
		% }
		% \subfigure{%
		% 	{\includegraphics[width=0.4\columnwidth]{fig/Household_labor_hours_dist_1}}%	
		% }
		
		% \subfigure{%
		% 	{\includegraphics[width=0.4\columnwidth]{fig/Household_norm_reward_dist_0}}%
		% }
		% \subfigure{%
		% 	{\includegraphics[width=0.4\columnwidth]{fig/Household_norm_reward_dist_1}}%
		
		% }
		
		\caption{Transitory impact of tax credits on households. See the increase in savings and consumption for all households.}\label{fig:tc_test}
	\end{figure*}

	\subsection{Transitory Impact of Tax Credits}\label{sec:tax_credit_study}
    {To assess the transitory impact of tax credits on the economic system, we first train agent policies in the absence of tax credits. The trained policies are then tested in the presence of fixed tax credits to observe changes in strategies.}
	% To assess the transitory impact of tax credits on the economic system, we simulate the test model by employing a pre-trained policy developed without considering tax credits. 
 {Hence, for this experiment, we learn policies for} three households, one firm, and the central bank, over a training horizon of three years so that $H=12$ quarters. To simulate households with different liquidities, recall that 
 high-liquidity households tend to prioritize saving over spending. In contrast, low-liquidity households lean towards increased spending. To reflect this tendency, we set the parameters for isoelastic utility of the three households as $\gamma_1 = 0.1, \mu_1 = 0.4, \gamma_2 = 0.2, \mu_2 = 0.2,\gamma_3 = 0.3, \mu_3 = 0.1 $ in order of decreased liquidity from Household 1 to Household 3. 
 % For the central bank, the weight in (\ref{reward_cb}) assigned to controlling the productivity target relative to the inflation target is set as $\lambda=1$. 
 Throughout this process, we set distinct learning rates for each entity:  $0.01$ for households, $0.005$ for the firm, and $0.001$ for the central bank. The discounted cumulative rewards during training are presented in Figure \ref{fig:tax_credit}. It shows that the low-liquidity Household 3 has a lower reward compared to the high-liquidity Household 1.
	
 	\begin{figure}[tb]
		\centering
		\includegraphics[width=\columnwidth]{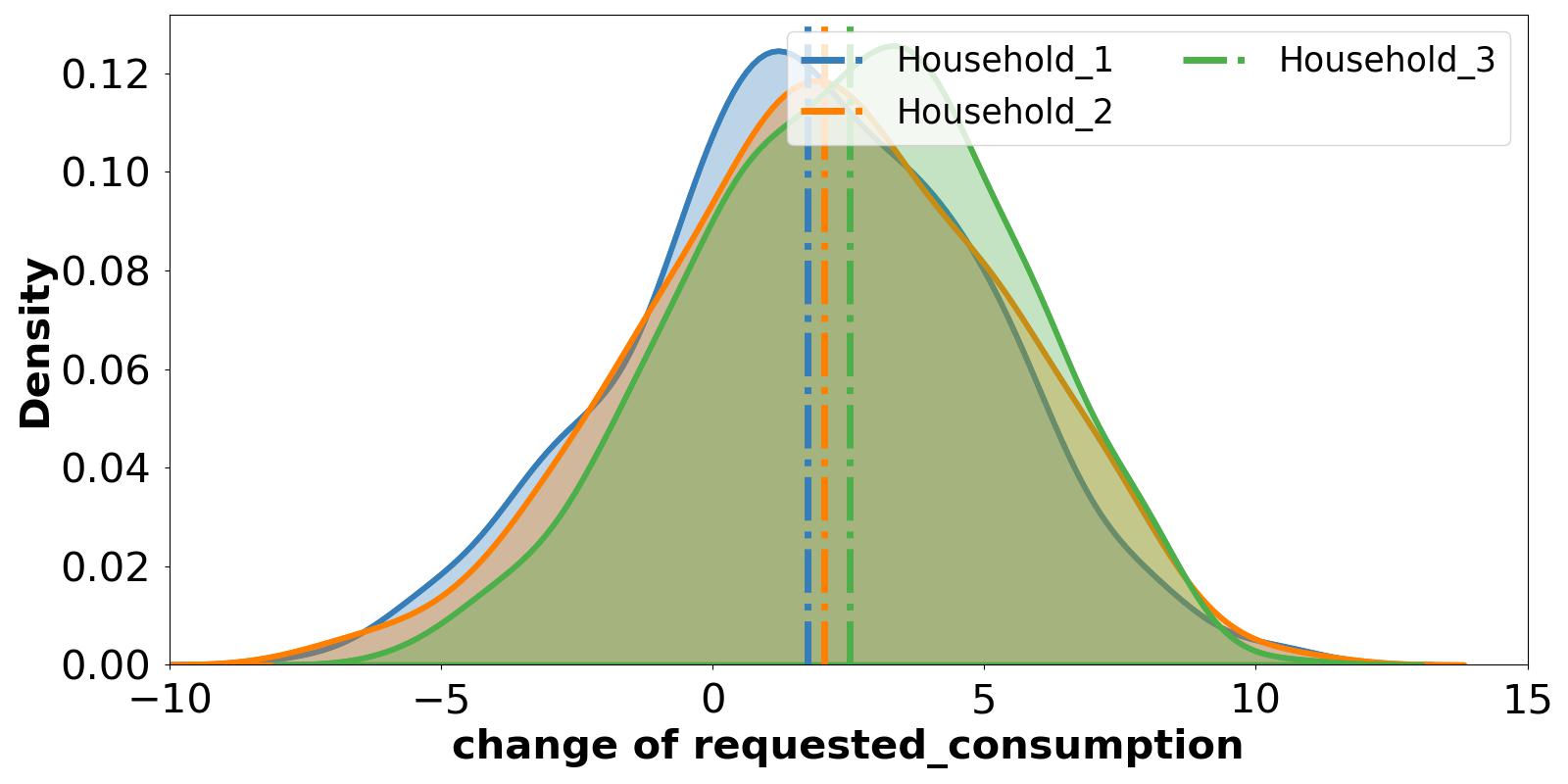}
		\caption{Change in requested consumption with tax credits. Household 3 with lowest liquidity sees the highest increase in consumption with tax credits.}\label{fig:change_tax_credit}
	\end{figure}
 {During the testing phase, we play out the trained policies in experiments without and with a tax credit of \$5000 per household per quarter. The distributions of household observables in test episodes are compared in Figure \ref{fig:tc_test}. }
	% During the testing phase, we compare the observation distributions of households without and with tax credit when using the trained policy presented in Figure \ref{fig:tax_credit}. For the second scenario, the government distributes each household $5000$ dollar tax credit at the beginning of each quarter. 
 % The simulation results are illustrated in Figure \ref{fig:tc_test}. 
 It shows that Household 3 has the least savings and highest requested consumption, yielding the lowest liquidity. The impact of tax credits on households exhibits increased savings and consumption. Moreover, Figure \ref{fig:change_tax_credit} illustrates the change in average requested consumption per quarter with tax credit from that without. It shows that low-liquidity households experience a more significant boost in their requested consumption following the receipt of tax credits than higher liquidity households.

 % \textcolor{blue}{We also see that the reduction in labor hours for H2 is higher than that for H1. Is it possible to get higher increase in consumption for H2?} \textcolor{red}{It may be possible if the increase in H2 consumption is smaller than the increase in H1 consumption.}
 % These results indicate that the introduction of tax credits has a positive impact on household consumption and savings. We now investigate the impact of tax credits on the source of household spending for low-liquidity and high-liquidity households.
 % , and study how to distribute the tax credits to households to improve social welfare.
	
	\begin{figure*}[t]
		\centering
		\includegraphics[width=\linewidth]{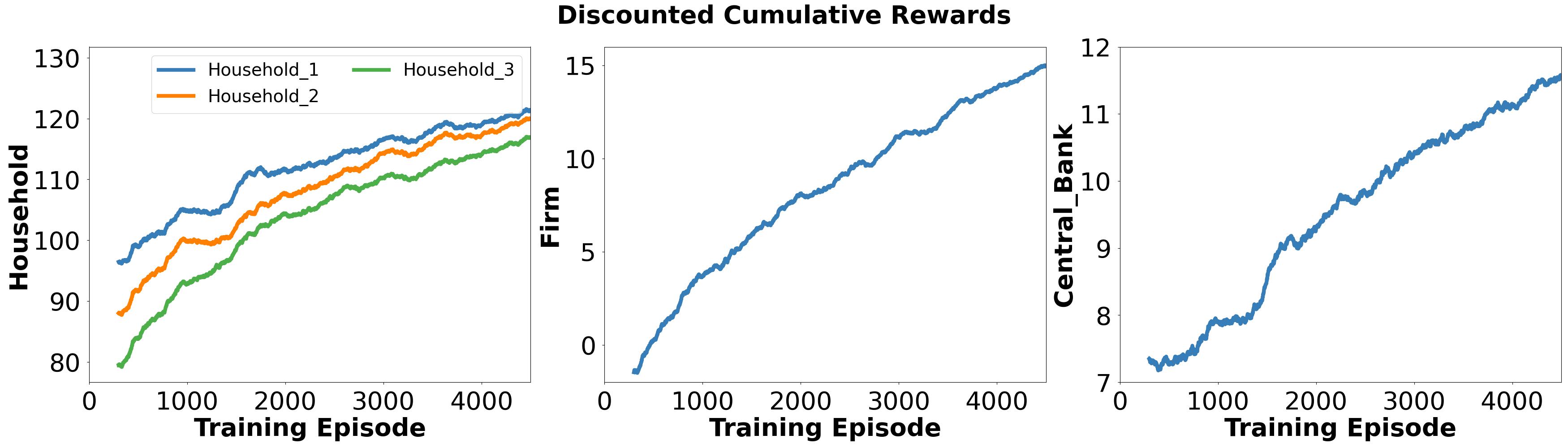}
		\caption{Training rewards with fixed tax credit distribution.}\label{fig:H_m}
	\end{figure*}
	
	\begin{figure}[t]
		\centering
		\includegraphics[width=\columnwidth]{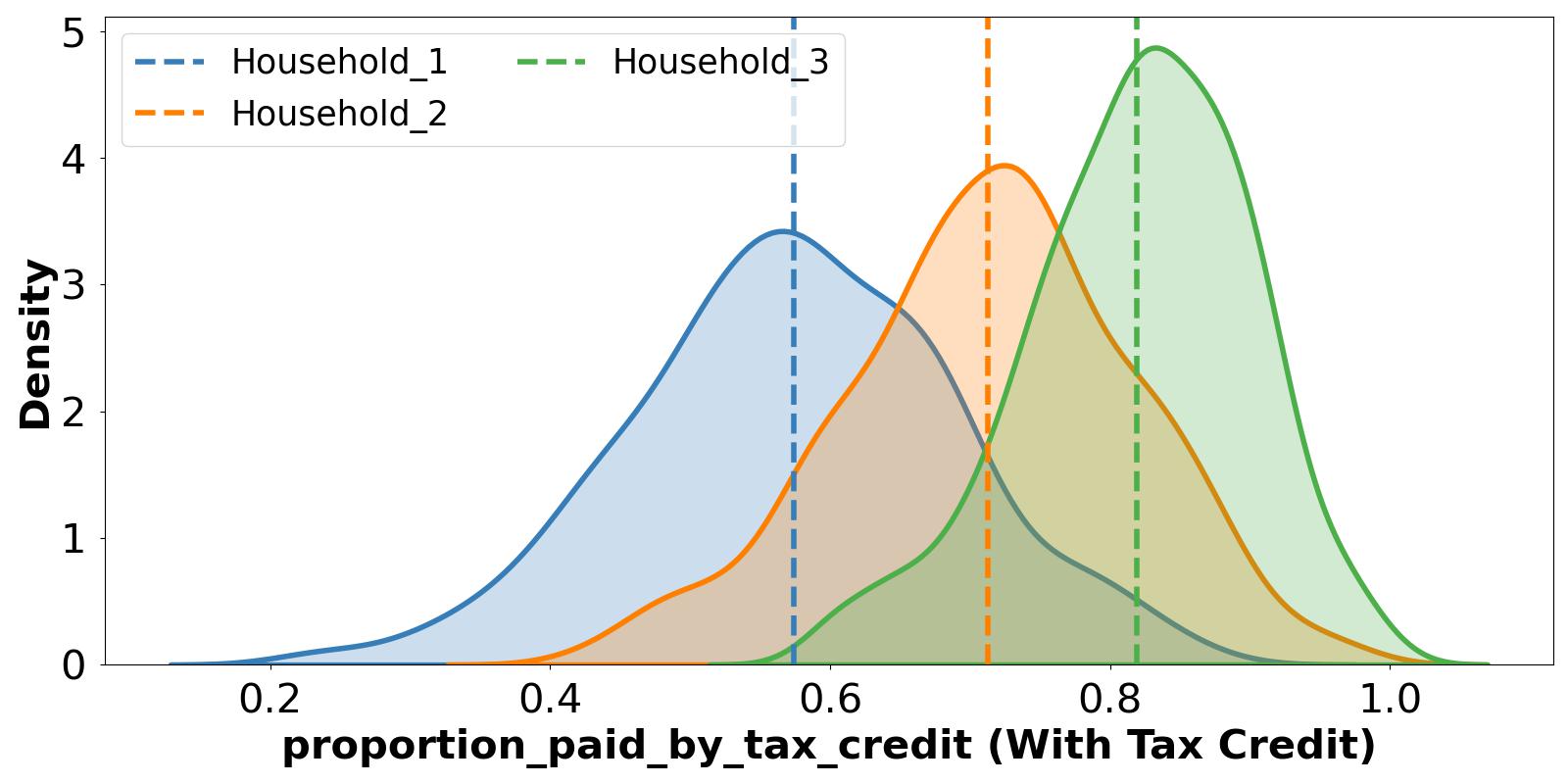}
		\caption{Spending patterns of fixed tax credits. Household 3 with lowest liquidity spends more of tax credit on consumption than others.}\label{fig:H_m_test}
	\end{figure}
	
	\begin{figure}[t]
		\centering
		\includegraphics[width=\columnwidth]{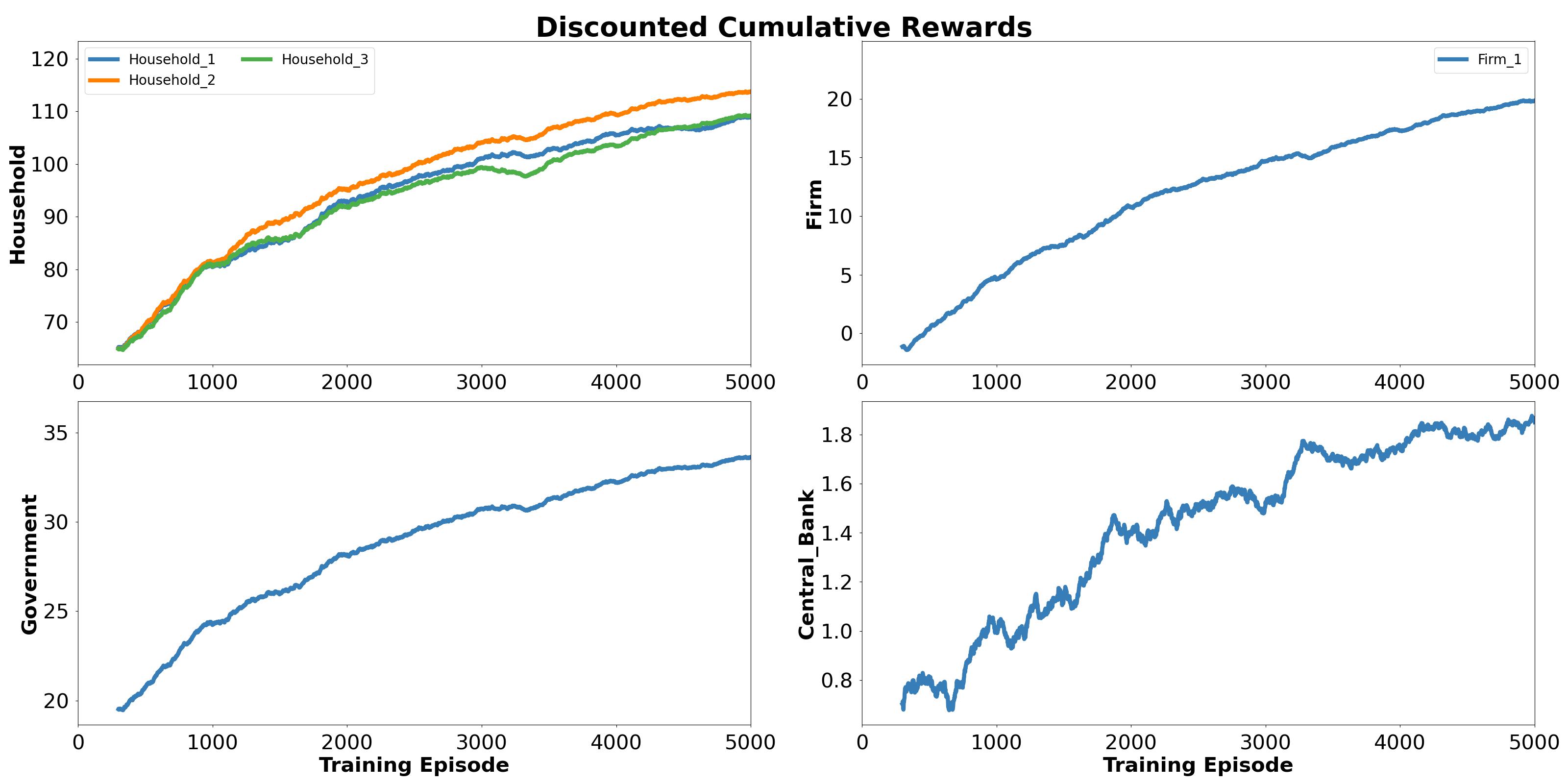}
		\caption{Training rewards with Government learning to distribute tax credits.}\label{fig:G_m}
	\end{figure}
 
	\subsection{Impact of Tax Credits on Household Spending Patterns}\label{sec:results_prop}
 Going a step further from studying the transitory impact of tax credits in the previous section, we now investigate the spending pattern of households following tax credit allocation. 
	% In addition to the modifications observed in households' behaviors and conditions as presented in the preceding subsection, our investigation then delves into the spending pattern of households following tax credit allocation. 
 {We employ the same agents as in Section \ref{sec:tax_credit_study} with the caveat that households have an additional action $\eta$ as described in Section \ref{subsec:3.1}. Additionally, all agents are now trained in the presence of a tax credit of \$5000 per household per quarter.}
 % In this section, we employ an economic model presented in Section \ref{sec:3}, encompassing three households, one firm, the central bank, and the government. The economic system setting is the same as Section \ref{sec:tax_credit_study}.
	% The policy is established through training three households, one firm, and the central bank. 
 The learning rates employed are $0.002$ for households, $0.005$ for the firm, and $0.005$ for the central bank. 
 % Throughout the training process, each household is allocated a quarterly tax credit of $5000$ dollars. 
 The training rewards for each agent are shown in Figure \ref{fig:H_m}. As before, Household 1 with the highest liquidity enjoys the highest reward among all households. 
	
	{To evaluate the household spending patterns learned in the presence of tax credits, we play out the learned policies in test episodes to collect observations.}
	% To evaluate the spending patterns resulting from tax credits within the economic framework, we test the proposed model with a pre-trained policy. 
 Specifically, we plot the distribution of the new action $\eta$ depicting how households utilize tax credits towards consumption in Figure \ref{fig:H_m_test}. The result demonstrates that low-liquidity households tend to expend a greater portion of their tax credits towards consumption spending than higher liquidity households. This observation is consistent with findings from the JP Morgan Chase report \cite{JPMC}. Importantly, notice that with a fixed tax credit distribution scenario, low-liquidity households continue to receive less reward compared to their high-liquidity counterparts. This emphasizes the need to intelligently allocating collected tax revenue to households, especially those with lower liquidity, towards enhancing overall household welfare. 
 % Further elaboration on the strategy of distributing tax credits will be presented in the following subsection.

	\subsection{Strategy of Distributing Tax Credits}\label{sec:results_gov}
	
	\begin{figure}[tb]
		\centering
		\includegraphics[width=\columnwidth]{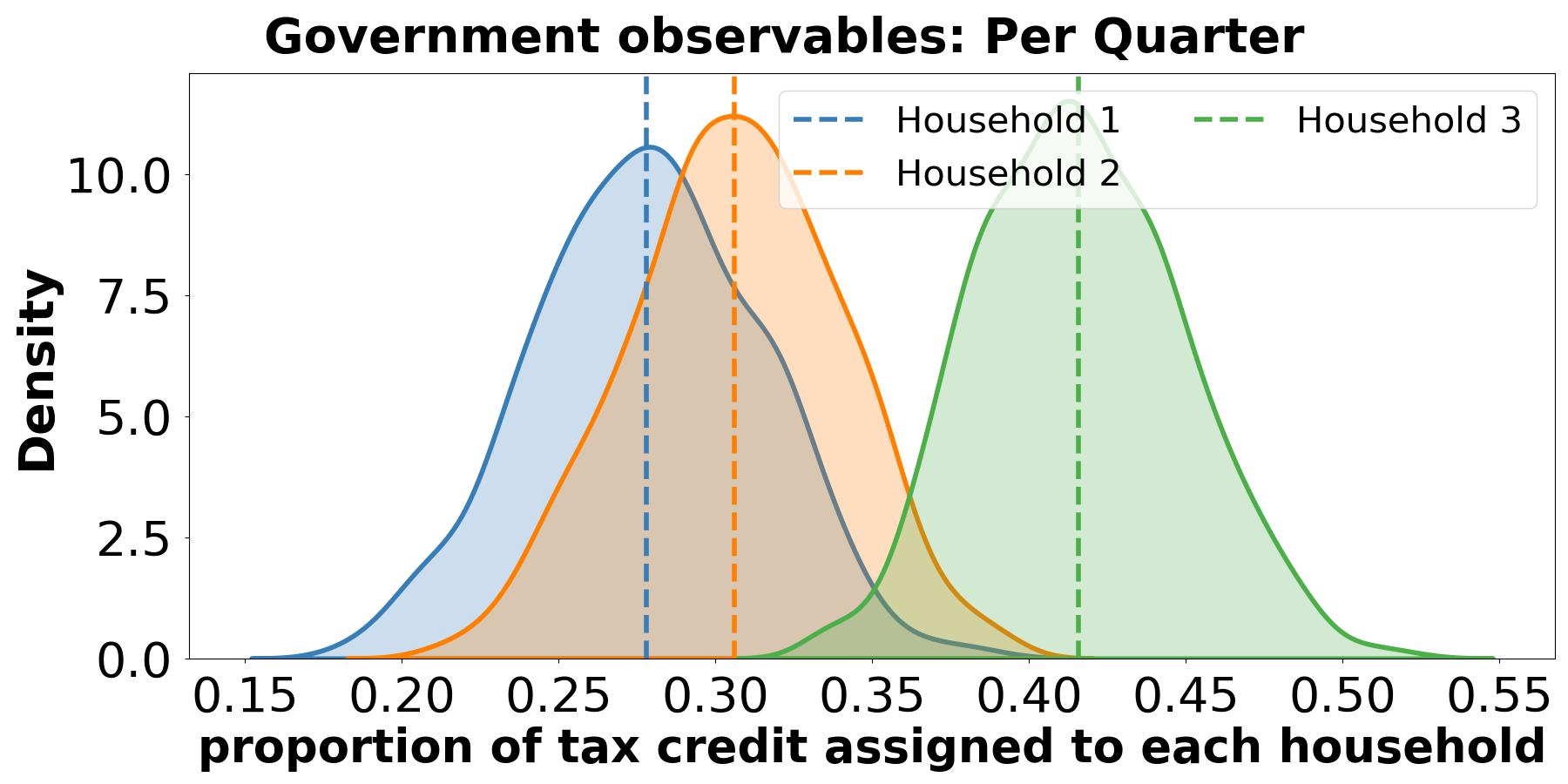}
		\caption{Government strategy for distributing tax credits. See that Household 3 with lowest liquidity receives a larger proportion of tax revenue as credits.}\label{fig:G_m_test}
	\end{figure}
 
	To improve the efficiency of tax credit distribution by the government, we integrate the government model proposed in Section \ref{sec:gov} into the economic system setting in Section \ref{sec:tax_credit_study}. 
 % This integration marks a significant advancement in our research, as it allows us to analyze the impact of government intervention on tax credit allocation strategies. By incorporating the government model into the training process, we create a more comprehensive simulation that mirrors real-world policy implementation. This approach enables us to assess not only the effectiveness of tax credit allocation methods but also the influence of government policies on economic agents' behaviors and decisions. Through this integrated analysis, we aim to provide valuable insights for policymakers, guiding them toward evidence-based decisions that can enhance the efficiency and fairness of tax credit distribution in the real economy.
We now learn policies for all agents in the economic system, with learning rates set at $0.002$ for households, $0.005$ for the firm, $0.005$ for the central bank, and $0.001$ for the government. The training rewards for each agent are represented in Figure \ref{fig:G_m}. Unlike the findings in Figure \ref{fig:H_m} with fixed tax credit assignments, the rewards among households appear more consistent in Figure \ref{fig:G_m}. This consistency in rewards signifies a more equitable distribution of resources among households, indicating the effectiveness of the integrated government model. 
% Furthermore, this approach not only balances the rewards across households but also promotes economic stability by ensuring that tax credits are allocated in a manner that aligns with both individual household needs and broader economic objectives. 
% By showcasing these results, our research underscores the importance of considering government intervention in tax credit allocation strategies, providing a foundation for future studies and policy implementations aimed at fostering a more just and stable economic environment.

	We test the trained agent policies to establish the efficacy of the tax credit distribution strategy. The simulation results illustrating the government's distribution of tax credits to households are shown in Figure \ref{fig:G_m_test}. The visualization shows the government's tendency to allocate greater tax credits to households with lower liquidity. This strategic approach stands to substantially enhance the financial well-being of low-liquidity households, thereby fostering a more equitable economic landscape.
	
	 % This integration marks a significant advancement in our research, as it allows us to analyze the impact of government intervention on tax credit allocation strategies. 
  By incorporating the government model into the training process, we create a more comprehensive simulation that mirrors real-world policy implementation. This approach enables us to assess not only the effectiveness of tax credit distribution methods but also the influence of government policies on economic agents' behaviors and decisions. Through this integrated analysis, we aim to provide valuable insights for policymakers, guiding them toward evidence-based decisions that can enhance the efficiency and fairness of tax credit distribution in the real economy.
  
	\section{Conclusion}
    We investigate the impact of tax credit allocation on the spending and saving behavior of diverse households in an economic system. We propose a multi-agent economic model encompassing heterogeneous households, a firm, central bank, and government. Each agent is equipped with reinforcement learning capabilities to adapt their strategies in response to others. By characterizing households by their liquidity (amount of savings versus consumption spending), we analyze their spending response to fixed and uniform tax credits.
    % To understand the spending behaviors exhibited by high-liquidity and low-liquidity households in response to tax credits, we introduce an innovative household model to capture the pattern of spending tax credits. 
    The outcomes of simulations distinctly highlight that households with lower liquidity tend to allocate a larger proportion of their tax credits toward consumption. This in turn affects their ability to save given necessary consumption spending with uniform credits. 
We subsequently propose a government strategy aimed at optimizing the distribution of tax credits to improve social welfare across households. Our simulation results demonstrate the effectiveness of this proposed strategy in ameliorating household inequalities, showing higher credit allocation to lower liquidity households.
% showing its potential to ameliorate inequalities among diverse households by distribtuinf tax cre. 
% Crucially, our proposed framework economic model establishes the foundation for an equitable economic framework, harmonizing with the complexity of the real world.

Crucially, our work paves the way for the advancement of multi-agent reinforcement learning within economic models. The potential for exploration and expansion is vast. Enhancing the model's complexity involves expanding the number of households and firms and developing efficient algorithms to train the multiple agent policies. 

\section*{Acknowledgments}
This paper was prepared for informational purposes in part by the Artificial Intelligence Research group of JPMorgan Chase \& Co. and its affiliates (``J.P. Morgan'') and is not a product of the Research Department of J.P. Morgan.  J.P. Morgan makes no representation and warranty whatsoever and disclaims all liability, for the completeness, accuracy or reliability of the information contained herein.  This document is not intended as investment research or investment advice, or a recommendation, offer or solicitation for the purchase or sale of any security, financial instrument, financial product or service, or to be used in any way for evaluating the merits of participating in any transaction, and shall not constitute a solicitation under any jurisdiction or to any person, if such solicitation under such jurisdiction or to such person would be unlawful.   

\bibliography{aaai24}

\end{document}